\documentclass[conference,9pt]{IEEEtran}
\IEEEoverridecommandlockouts
\usepackage{cite}
\usepackage{amsmath,amssymb,amsfonts}
\usepackage{algorithmic}
\usepackage{graphicx}
\usepackage{textcomp}
\usepackage{xcolor}
\usepackage[utf8]{inputenc}
\usepackage{svg}
\usepackage{subcaption}
\usepackage{graphicx}
\usepackage{blindtext}
\usepackage{hyperref}
\usepackage{enumitem}
\usepackage{booktabs}
\definecolor{softgreen}{RGB}{102, 194, 165} % Soft green
\definecolor{softred}{RGB}{252, 141, 98}    % Soft red
\usepackage[english]{babel}

\def\BibTeX{{\rm B\kern-.05em{\sc i\kern-.025em b}\kern-.08em
    T\kern-.1667em\lower.7ex\hbox{E}\kern-.125emX}}

\usepackage{titlesec}
\titlespacing*{\section} {0pt}{1ex}{0.5ex}
\titlespacing*{\subsection} {0pt}{1ex}{0.5ex}

\begin{document}

\title{Designing with Deception: ML- and Covert Gate-Enhanced Camouflaging to Thwart IC Reverse Engineering}

% \author{\IEEEauthorblockN{Removed for double-blind review.}
% \IEEEauthorblockA{\textit{} \\
% %\textit{}\\
% %\\
% {}
% }}

\author{\IEEEauthorblockN{Junling Fan, David Koblah, Domenic Forte}
\IEEEauthorblockA{\textit{Department of Electrical and Computer Engineering} \\
\textit{University of Florida}, Gainesville, USA \\
{\{fan.j, dkoblah\}@ufl.edu, dforte@ece.ufl.edu}
}}

% \and
% \IEEEauthorblockN{2\textsuperscript{nd} David Koblah}
% \IEEEauthorblockA{\textit{ECE Department} \\
% \textit{University of Florida}\\
% Gainesville, USA \\
% dkoblah@ufl.edu}
% \and
% \IEEEauthorblockN{3\textsuperscript{rd} Domenic Forte}
% \IEEEauthorblockA{\textit{ECE Department} \\
% \textit{University of Florida}\\
% Gainesville, USA \\
% dforte@ece.ufl.edu}

\maketitle

\begin{abstract} Integrated circuits (ICs) are essential to electronic systems, yet they face significant risks from physical reverse engineering (RE) attacks that compromise intellectual property (IP) and overall system security. While IC camouflaging has emerged to mitigate these risks, existing approaches largely focus on localized gate modifications, neglecting comprehensive deception strategies. To address this gap, we present a machine learning (ML)-driven methodology -- IP Camouflage -- that integrates cryptic and mimetic deception principles to enhance IC security against RE.
Our approach leverages a novel And-Inverter Graph Variational Autoencoder (AIG-VAE) to encode circuit representations, enabling dual-layered camouflage through functional preservation and appearance mimicry. By introducing new variants of covert gates -- Fake Inverters, Fake Buffers, and Universal Transmitters -- our methodology achieves robust protection by obscuring circuit functionality while presenting misleading appearances. Experimental results demonstrate the effectiveness of our strategy in maintaining circuit functionality while achieving strong resistance to SAT-based attacks with low structural overhead.
Additionally, we validate the robustness of our method against advanced artificial intelligence (AI)-based RE attacks. 
%By bridging the gap in mimetic deception, our work sets a new standard for IC camouflage, advancing the application of deception principles to protect critical systems from adversarial threats. 
\end{abstract}


\section{Introduction}

As integrated circuits (ICs) continue to serve as the backbone of modern electronic systems, the threat of reverse engineering (RE) has become increasingly consequential. Physical RE enables adversaries to extract sensitive information from hardware designs, including intellectual property (IP), underlying algorithms, and security features. By physically probing, delayering and imaging, and analyzing the results, attackers can reconstruct an IC's gate-level representation, enabling replication, cloning, vulnerability analysis, and/or tampering~\cite{RE2016}. This poses substantial risks, particularly in critical domains such as defense, healthcare, and finance, where IP theft or hardware tampering could have catastrophic consequences to national security, public health and economic security, respectively. With the continuous advancement of RE tools and techniques, safeguarding ICs from these threats has become increasingly challenging \cite{torrance2011state,JinSurvey2019}.

IC camouflaging has emerged as a technique to protect hardware IPs from RE attacks. This approach leverages fabrication-based technologies to intentionally obscure the true functionality of the circuit, deliberately confusing adversaries~\cite{cocchi2014circuit, rajendran2013security}. Camouflaged sections are designed to mislead attackers during the IC RE process, resulting in incorrect or incomplete reconstructions. While this may be the intention, current methods primarily focus on \textit{localized} gate layouts or interconnections, neglecting a holistic camouflage of the overall circuit's functionality. This narrow focus limits their ability to achieve broader, system-level deception, which could significantly hinder adversarial efforts during RE in real-world applications.

Cyber deception offers a proactive and versatile defense strategy for addressing these limitations. 
%By misleading and confusing attackers, cyber deception enhances system resilience against adversarial efforts. 
The taxonomy proposed by Pawlick et al.~\cite{CD2019} categorizes cyber deception into six distinct strategies: perturbation, moving target defense (MTD), obfuscation, mixing, honey-x, and attacker engagement. These are classified broadly into two categories: \textit{cryptic} deception, which obscures the existence or nature of ``assets'', and \textit{mimetic} deception, which simulates realistic but misleading appearances~\cite{CD2019}. According to this definition, existing camouflage strategies against IC RE focus on cryptic deception while mimetic deception %-- emphasizing imitation, and false beliefs -- 
remains unexplored.

Our work addresses this through the following contributions: \begin{itemize}[wide, labelwidth=!, labelindent=0pt]
\item \textbf{Filling the Mimetic Deception Gap:} We introduce mimetic deception, complementing cryptic deception by simultaneously hiding the true functionality of circuits and presenting a misleading outward appearance. \autoref{ToyExample} shows a toy example where a camouflaged Half Adder (HA) looks similar to a Full Adder (FA) in appearance. We also create new variants of covert gates~\cite{shakya_covert_2019} which would appear as inverters, buffers, etc. under SEM but may function differently.

\begin{figure}[t]
    \centering
    \includegraphics[width=0.5\textwidth]{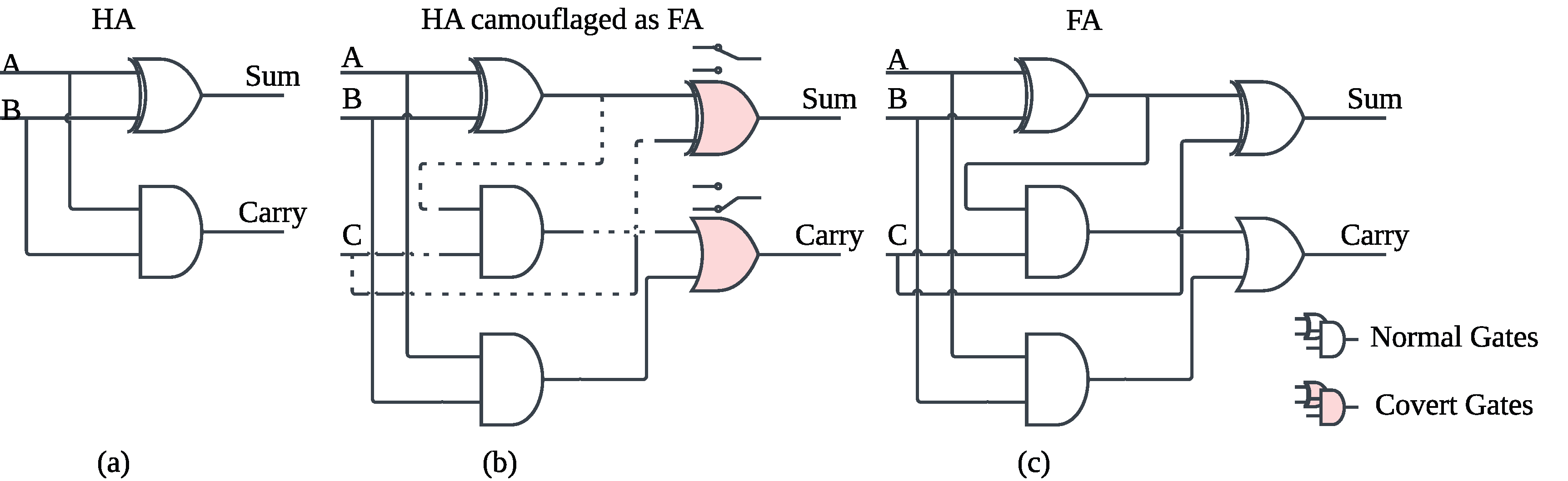}
    \caption{Illustration of mimetic deception for: (a) a Half Adder (HA), (b) an HA camouflaged deceptively to appear as a Full Adder (FA) using covert gates~\cite{shakya_covert_2019}, and (c) a standard FA. The pink gates are covert gates that appear as the logic gate under SEM imaging but are actually buffers or inverters that only operate on the non-dashed input signal. Dashed lines are dummy inputs unbeknownst to attackers.}
    \label{ToyExample}

\vspace{-3ex}
    
\end{figure}

\item \textbf{Machine Learning-Based Implementation:} We employ an ML-based strategy using a variational autoencoder (VAE) model to achieve ``dual-layered'' camouflage with and-inverter graphs (AIGs); that is, it ensures efficient and systematic implementation of both cryptic and mimetic deception, overcoming the limitations of traditional approaches. Note that the proposed AIG-VAE is also a unique contribution which can be used in other EDA/CAD applications.

\item \textbf{Demonstrated Robustness Against AI-Enhanced RE Tools:} Through experiments, we demonstrate how our strategy effectively thwarts advanced RE methods, including those leveraging AI, underscoring its practical security improvements. %We also propose scoring metrics to capture the effectiveness of AIG-VAE-based camouflaging.

% \item \textbf{New CG}

\end{itemize}

% By integrating these, similar approaches are used to represent circuits, but they are not typically employed for generation, or lack motivation for generation using VAE-like models, as the exploration of the latent space is not explicit.
%Ultimately, our work advances the application of cyber deception principles in hardware security, providing a novel, robust, and scalable methodology to enhance the resilience of ICs against RE attacks.

The remainder of this paper is organized as follows: Section~\ref{sec:attack_models} offers background on threats against ICs and discusses how they may be impacted by the proposed camouflaging. % approach and vice versa.  
In Section~\ref{sec:related}, we provide background on the main sub-components of the proposed approach: covert gates, VAEs, and cyber deception. The section concludes with a few motivational examples, highlighting the vast potential of hardware security that arises from combining IC camouflaging with deception principles. The overall methodology (IP Camouflage) and main contributions are discussed in Section~\ref{sec:method}. Experimental results that demonstrate the effectiveness of the AIG-VAE and IC camouflage are provided in Section~\ref{sec:results}. %These results are re-connected to the threats in Section~\ref{sec:attack_models}. 
Finally, Section~\ref{sec:conclusion} concludes the paper and highlights directions for future research.

\section{Attacks and Threat Model}\label{sec:attack_models}
The landscape of hardware security has rapidly evolved in the last few decades, driven by the changing nature of attacks and the corresponding development of countermeasures. 
The constant theme in this field is that there is no one-size-fits-all solution for all adversaries.
% In this work, we focus on the attack models outlined below.
However, our methodology improves IP security against a few distinct physical attacks.
    \subsection{Reverse Engineering}
    Reverse Engineering (RE) is a process that involves systematically analyzing electronic devices to extract design information at the chip, board, or system levels. It involves dissecting hardware designs to uncover vulnerabilities or proprietary information~\cite{rekoff1985reverse}.
    %The ultimate goal of RE is to create a product that replicates the utility and functionality of the original. %As shown in the \autoref{RE}, 
    A typical process of IC RE includes:
\begin{enumerate}[wide, labelwidth=!, labelindent=0pt]
    \item Remove the chip packaging %(\textbf{Decapsulation}) 
    to expose the die for further analysis.
    \item Sequentially remove individual chip layers %(\textbf{Deprocessing}) 
    for examination.
    \item Capture high-resolution images of the exposed layers. %(\textbf{Imaging}).
    \item Annotate the captured images and extract the circuit's netlist %(\textbf{Annotation and Netlist Extraction}) 
    using hardware assurance tools.
\end{enumerate}

Adversaries can use RE for cloning, counterfeiting, and IP theft. Techniques like SEM imaging, dry or wet etching, and annotation tools are employed for accurate extraction and analysis of device designs~\cite{torrance2011state, RE2016}.
    Given the complexity of modern hardware, the process requires significant resources.
    Although it has utility in security and assurance applications, such as hardware Trojan detection~\cite{bao2015reverse,rajendran2021novel}, it is viewed as an attack in this paper.

    \subsection{SAT-based Attack}
    The SAT problem asks whether a given Boolean formula \( F(x_1, x_2, x_3, \ldots, x_n) \) is satisfiable, i.e., whether there exists an assignment of variables \( x_i \,\forall \,i\) that makes \( F \) evaluate to true.
    While SAT is NP-complete, modern SAT solvers are highly efficient and have been successfully applied in various domains such as equivalence checking, formal verification, and automatic test pattern generation (ATPG)~\cite{yasin2020sat, el2019sat}.
     By leveraging the power of SAT solvers~\cite{wang2024cimsat, zhou2017cycsat, shen2019sigattack},  circuit obfuscation, including logic locking and camouflaged circuits has been broken. Such attacks assume that a locked/camouflaged netlist is available along with a functional/unlocked chip (oracle). Then, the following iterative process occurs:
    \begin{enumerate}[wide, labelwidth=!, labelindent=0pt]
        \item Identify a distinguishing input pattern (DIP) using the SAT solver. 
        This is an input that differentiates between the outputs produced by two or more candidate keys\footnote{For IC camouflaging, the netlist can be translated to one with keys where the correct key bits identify the correct function of the camouflaged cells.}.
        \item Apply the DIP to oracle (unlocked chip) and observe the correct output response.
        \item Rule out incorrect key assignments by adding constraints to the SAT solver that eliminate keys producing outputs inconsistent with the oracle's response.
        \item Repeat the above steps until no further DIPs can be found. At this point, the correct key (de-camouflaged design) is uniquely identified.
    \end{enumerate}
The basis of the proposed approach is covert gates (see Section~\ref{subsec:covert}) which SAT and VLSI-test based attacks do no scale well against~\cite{shakya_covert_2019}. Further, our ground-up approach that extends camouflaging to the IP level should also prevent the success of these attacks.

    \subsection{Threat Model in This Work}
This work addresses a threat model where an adversary, such as an IC reverse engineer, seeks to exploit ICs to compromise their security and functionality. The adversary’s primary objectives are: 

\begin{enumerate}[wide, labelwidth=!, labelindent=0pt]
    \item \textbf{Uncovering IC Functionality}: Using reverse engineering techniques to extract critical design information from security-focused IP, such as intrusion detection systems or encryption modules, to bypass or exploit their functionality. Further, the IP can also be stolen, replicated, and counterfeited.
    \item \textbf{Tampering with Sensitive Signals}: Using the information from \# 1 to target critical components or operations through physical attacks, such as laser fault injection (LFI) or focused ion beam (FIB), to disrupt normal functionality or manipulate sensitive signals.
\end{enumerate}
We assume that the adversary has access to physical RE equipment, such as SEM imaging, but optical side channels~\cite{rahman2020key} are counteracted by sensing techniques, e.g.,~\cite{monfared2024laserescape}. Fault injection and non-invasive side channel analysis, which are underexplored in de-obfuscation, are considered out of scope and left for future work.

\section{Related Works} \label{sec:related}
    \subsection{IC Camouflaging and Covert Gates} \label{subsec:covert}
    IC camouflaging can generally be classified into two categories~\cite{shakya_covert_2019}: gate and interconnect camouflaging. Gate camouflaging focuses on replacing individual logic gates within the circuit with ones that could implement multiple functions depending on fabrication-related secrets (e.g., dummy contacts~\cite{rajendran2013security}, threshold voltages~\cite{erbagci2016secure}, or semiconductor doping~\cite{malik2015development}). This approach effectively disrupts the ability of the attacker to identify the specific logic gates from SEM images. On the other hand, interconnect camouflaging~\cite{yu2017incremental} involves altering the connections between gates, making it difficult for an adversary to accurately map out the circuit's wiring or logic flow during IC RE.

    The state-of-the-art (SotA) in IC camouflaging is covert gates\footnote{To our knowledge, circuits implemented with covert gates have never been successfully broken in the literature. Thus we consider it to be the SotA.}, which effectively combine gate and interconnect variants through the use of always-on and always-off transistors~\cite{shakya_covert_2019}. The designer can create logic gates with different functions as well as circuits with dummy inputs (dummy connections) by altering transistor states from normal to always-on and always-off. Further, since the transistor states are not visible to attackers under SEM, the covert gates have the same appearance as standard CMOS gates, allowing them to achieve higher scalability against SAT and VLSI test-based attacks compared to other camouflaged gates. Given these advantages, we adopt the covert gate methodology as part of our camouflaging strategy.
    
    \subsection{Variational Autoencoder (VAE)} \label{subsec:VAE}
    A variational autoencoder (VAE) \cite{VAE} is a generative model that encodes data into a latent space and reconstructs it, introducing a probabilistic framework for generating new samples. In a VAE, the encoder maps the input \( x \) to a latent variable \( z \) by learning a probability distribution \(z \sim q(z|x) = \mathcal{N}(\mu(x), \sigma(x)^2).\)

The decoder reconstructs \( x \) from \( z \) by generating \( p(x|z) \)\(
    \hat{x} \sim p(x|z).
\)
To ensure meaningful interpolation, the latent space is regularized with a Gaussian prior \( p(z) = \mathcal{N}(0, I) \), where \(I\) refers to the identity matrix.
The VAE’s objective combines two loss terms:  \textit{reconstruction loss}, which measures how well \( x \) is reconstructed from \( z \):\(
    \mathcal{L}_{\text{recon}} = -\mathbb{E}_{q(z|x)}[\log p(x|z)],
    \label{eq:recon_loss}
\)
and \textit{KL divergence loss}, which aligns \( q(z|x) \) with the prior \( p(z) \):
\(
    \mathcal{L}_{\text{KL}} = D_{\text{KL}}(q(z|x) \parallel p(z)),
    \label{eq:kl_divergence}
\)where $\parallel$ denotes the divergence between two probability distributions.
The total loss function for the VAE is
\begin{equation}
    \mathcal{L}_{\text{VAE}} = \mathcal{L}_{\text{recon}} + \mathcal{L}_{\text{KL}}.
    \label{eq:vae_loss}
\end{equation}

\begin{figure}[t]
    \centering
    \includegraphics[width=0.6\linewidth]{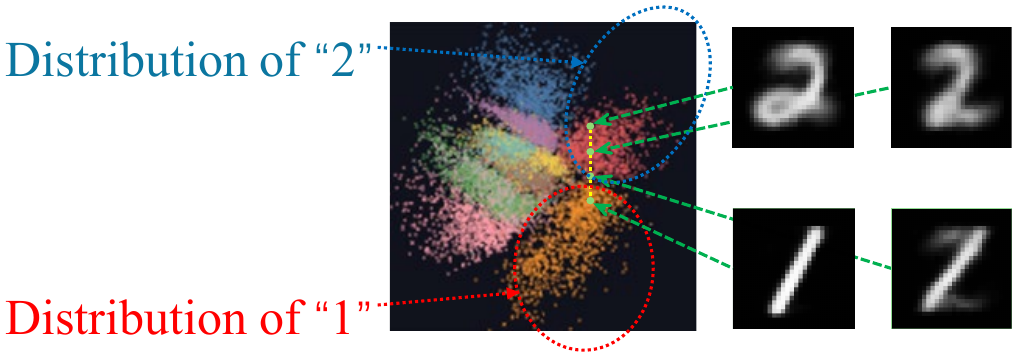} % Adjust the file path if needed
    \caption{
        Latent space distribution of the MNIST dataset~\cite{MNIST}. Using~\cite{VAEVisual}, a visualization shows how interpolation in the latent space can result in samples that mix features from different inputs.
    }
    \label{fig:vae_mnist}

    \vspace{-3ex}
    
\end{figure}

Machine learning models based on VAE enable interpolation between two latent codes from 2 different samples, allowing the creation of a new sample with a balanced feature between two samples as \autoref{fig:vae_mnist} shows. VAEs and such interpolation are the basis of our camouflaging approach where the two interpolated samples are the circuit's desired function and its desired appearance. 

\subsection{Cyber Deception: Cryptic and Mimetic Strategies}

Cyber deception represents a proactive and innovative defense mechanism designed to counter cyber threats by misleading attackers, complicating their operations, and enhancing overall system security. Unlike traditional defenses that primarily focus on detection and prevention, cyber deception actively manipulates the attack surface to confuse, disorient, and frustrate adversaries. This strategic approach is rooted in the historical principle of deception used in warfare and espionage but has been adapted to address the complex challenges of cybersecurity in the digital era \cite{CD2019, CD2024}.

\vspace{0.5ex}

\noindent \textbf{Cryptic Deception.} Cryptic deception aims to obscure the existence or true nature of assets, thereby \textit{hindering attackers from identifying their targets}. By hiding critical information, cryptic methods effectively render the attack surface ambiguous, complicating an adversary’s reconnaissance efforts and reducing the likelihood of a successful breach. A prominent example of cryptic deception is the use of moving target defenses (MTDs), which dynamically alter system configurations to create uncertainty and unpredictability. This approach forces attackers to expend significant time and resources without guaranteeing success. MTD strategies have been effectively applied to hardware security, as demonstrated in \cite{zhang2018thwarting,monfared2024randohm,monfared2024laserescape}, where hardware configurations are reconfigured at run time to protect designs and/or on-chip assets from hardware Trojans, impedance-related side-channel attacks, and optical probing.

\vspace{0.5ex}

\noindent \textbf{Mimetic Deception.} Mimetic deception, on the other hand, focuses on imitation to present misleading but realistic appearances. This approach seeks to mislead attackers by simulating legitimate system behaviors or mimicking the characteristics of valuable targets. Mimetic techniques include the deployment of honeypots and decoys, which imitate operational systems to \textit{lure attackers away from critical assets}. These methods %not only divert malicious activities but 
also serve as intelligence-gathering tools, enabling defenders to study attackers’ tactics, techniques, and procedures (TTPs) in real time \cite{CD2019, CD2024}.

\vspace{0.5ex}

\noindent \textbf{Relevance to Anti-RE and Hardware Security.} Our proposed methodology aligns with the principles of both cryptic and mimetic deception. By leveraging techniques such as functional preservation and appearance mimicry, we achieve a dual-layered camouflaging strategy. The functional integrity of an IP is preserved, ensuring operational reliability, while the appearance is camouflaged to resemble alternate IPs. This hybrid approach effectively misleads RE adversaries, aligning with cryptic principles by concealing the true logic of circuits and mimetic principles by presenting a deceptive outward appearance.
This strategy %not only enhances the resilience of systems against RE but 
also highlights the evolving role of cyber deception in modern security paradigms. By combining cryptic and mimetic deception, our work contributes to the broader landscape, % of cyber deception, 
demonstrating applicability to IP anti-theft as well as hardware security. Examples of promising new capabilities enabled by this approach include:
\begin{enumerate}[wide, labelwidth=!, labelindent=0pt]
\item \textbf{Misdirection to Hide Real On-chip Assets.} Consider an intrusion detection and prevention IP that monitors network traffic and detects potential security threats. Through IC RE, an attacker can gain an understanding of how this IP works, exploiting the information to evade detection or inject false alarms in subsequent attacks on products that use it. Our dual-layered approach could make this security-focused IP look like something innocuous such as a USB or memory controller, making it seem less  valuable as an attack target.
\item \textbf{Creation of On-chip Decoys to Attract Physical Attacks.} Consider an IP that detects LFI. Using IC RE, an attacker could identify the locations of this IP within the chip and thus places to avoid aiming the laser. Our dual-layered approach could make this security-focused IP appear to be a target of interest such as a cryptographic core. When an attacker tries to attack the decoy core, it would instead trigger data zeroization or chip self-destruction. 
\end{enumerate}
%Additionally, similar to MTD approaches in hardware security \cite{DavidMTD}, our work incorporates adaptability and unpredictability, which are essential features of effective cyber deception strategies.
\textit{The deception concept combined with anti-RE is unique to this paper and opens up many promising directions in hardware security.}

\section{Proposed Methodology: IP Camouflage} \label{sec:method}
    \subsection{Overview}
    \label{subsec:MethodOverview}
   Existing camouflaging approaches are localized, primarily residing at the gate-level where a gate's function can be one of several functions that the attacker needs to determine. Here, we propose \textit{IP Camouflage} to create a camouflaged IP that preserves an IP's function while mimicking the appearance of a different IP, misleading RE adversaries as described above. We refer to the IP with the desired functionality as the ``functional circuit''. % which serves as the prototype for the intended operation of the camouflaged design. 
   The IP whose appearance we want the product to resemble is called the ``appearance circuit''. The process begins by converting both circuits into And-Inverter Graphs (AIGs), which are then encoded into latent representations using the encoder of our Variational Autoencoder (VAE)-based model, the \textit{AIG-VAE} (described in \autoref{subsec:Proposed Machine Learning Model: AIG-VAE}). 

    The encoder generates probabilistic distributions for each circuit, represented by a mean (\( \mu \)) and standard deviation (\( \sigma \)) that follow a normal distribution. From these distributions, two latent space codes are sampled: \( \mathbb{Z}_F \), representing the functional circuit, and \( \mathbb{Z}_A \), representing the appearance circuit. These codes are interpolated using a proportion factor \( p \), resulting in an intermediate latent code \( \mathbb{Z}_{G'} \), which combines the characteristics of the functional circuit and the appearance circuit: \(\mathbb{Z}_{G'} = (1 - p) \mathbb{Z}_F + p \mathbb{Z}_A\).

\begin{figure}[t]
    \centering
    \includegraphics[width=0.5\textwidth]{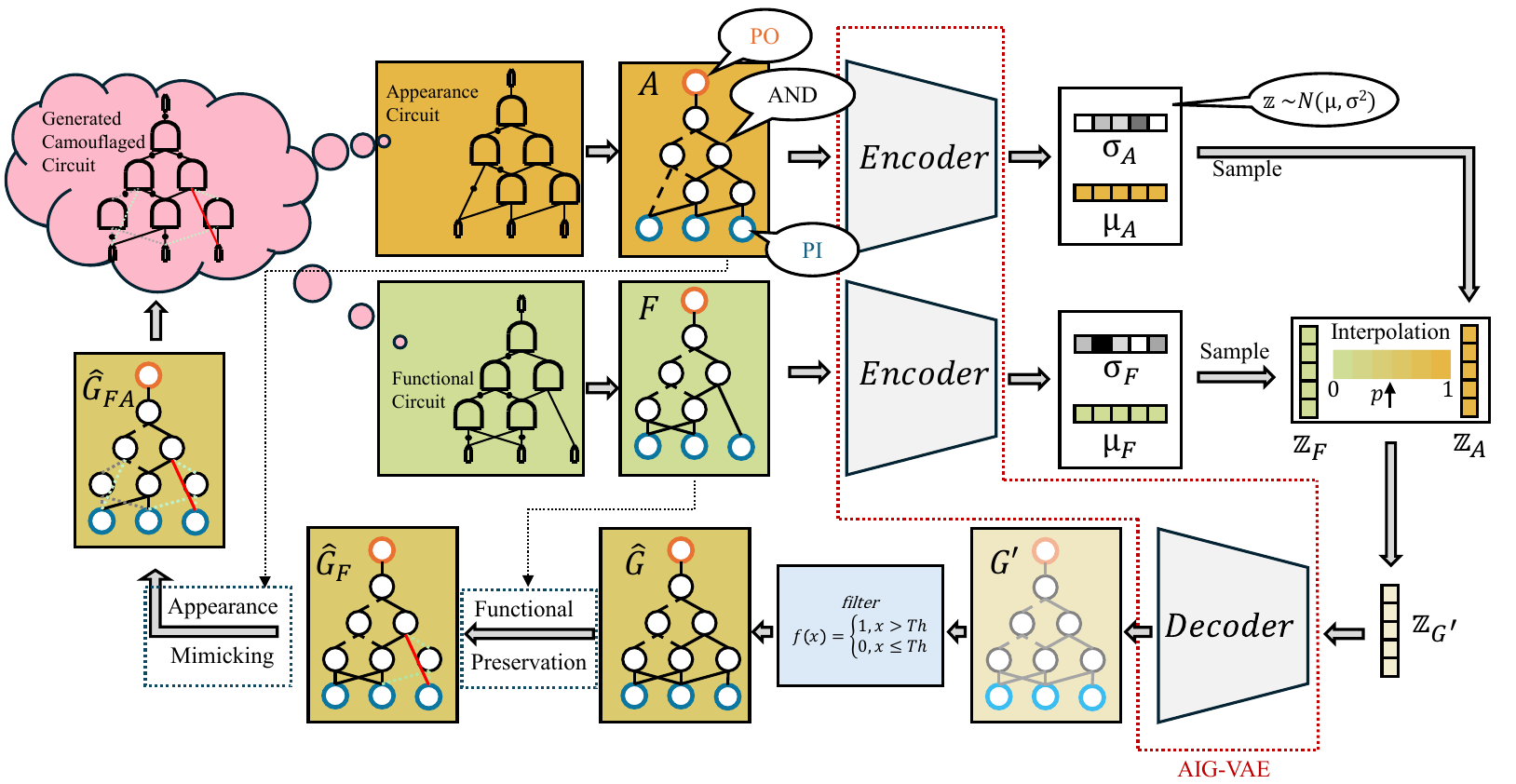}
    \caption{High-level Diagram of IP Camouflage. A functional circuit $F$ and appearance circuit $A$ are taken as inputs. After AIG-VAE and post-processing fixes, the final output is a camouflaged circuit \( \hat{G}_{FA} \) in AIG format that matches $F$'s function but appears as $A$.}
    \label{Overview}
    
    \vspace{-3ex}

\end{figure}
The decoder then reconstructs the circuit \( G' \) from \( \mathbb{Z}_{G'} \), followed by a threshold-based filter to refine the output, ensuring clear and interpretable connections:
\begin{equation}
filter(x, Th) = 
\begin{cases} 
0, & x \leq Th \\
1, & x > Th 
\end{cases}
\label{filter}
\end{equation}
The generated circuit \( \hat{G} \) is further refined with a \textit{Functional Preservation} to align its function with \( F \), producing \( \hat{G}_F \). After this, \textit{Appearance Mimicking} is applied to ensure the produced circuit resembles \( A \). The final result, denoted as \( \hat{G}_{FA} \), achieves both functional alignment with the functional circuit and visual similarity to the appearance circuit. The process is shown in \autoref{Overview}.

\vspace{0.5ex}

\noindent \textbf{Novelty.} Unlike existing approaches that focus solely on localized camouflage, IP Camouflage leverages a machine learning model (AIG-VAE) to achieve holistic, system-level camouflage and deception. Note, however, that camouflaged gates are still utilized and present in our circuits.  To be more specific, by integrating newly designed covert gates -- Fake Inverters (FI), Fake Buffers (FB), and Universal Transmitters (UT), introduced in \autoref{FBFIUT} -- during Functional Preservation and Appearance Mimicking, we enhance the robustness of the camouflage against adversaries. The subsequent sections provide detailed explanations of these processes.

    \subsection{Machine Learning Model: AIG-VAE}\label{subsec:Proposed Machine Learning Model: AIG-VAE}

To effectively camouflage a circuit, our model must first  understand the circuit’s structure and function. Previous works, such as FGNN~\cite{FGNN} and ABGNN \cite{ABGNN}, excel at learning functional representations of circuits by using asynchronous message-passing mechanisms, allowing the model to interpret the intricate relationships within circuit netlists. However, these models lack the generative capabilities needed to produce new circuits. This is where models like VGAE and D-VAE \cite{VariationalGraphEncoder,zhang2019dvaevariationalautoencoderdirected} are important. These VAE-based models enable graph generation, allowing circuits to be generated from informative embeddings derived from circuit representation learning. 

\begin{figure}[t]
    \centering
    \includegraphics[width=0.5\textwidth]{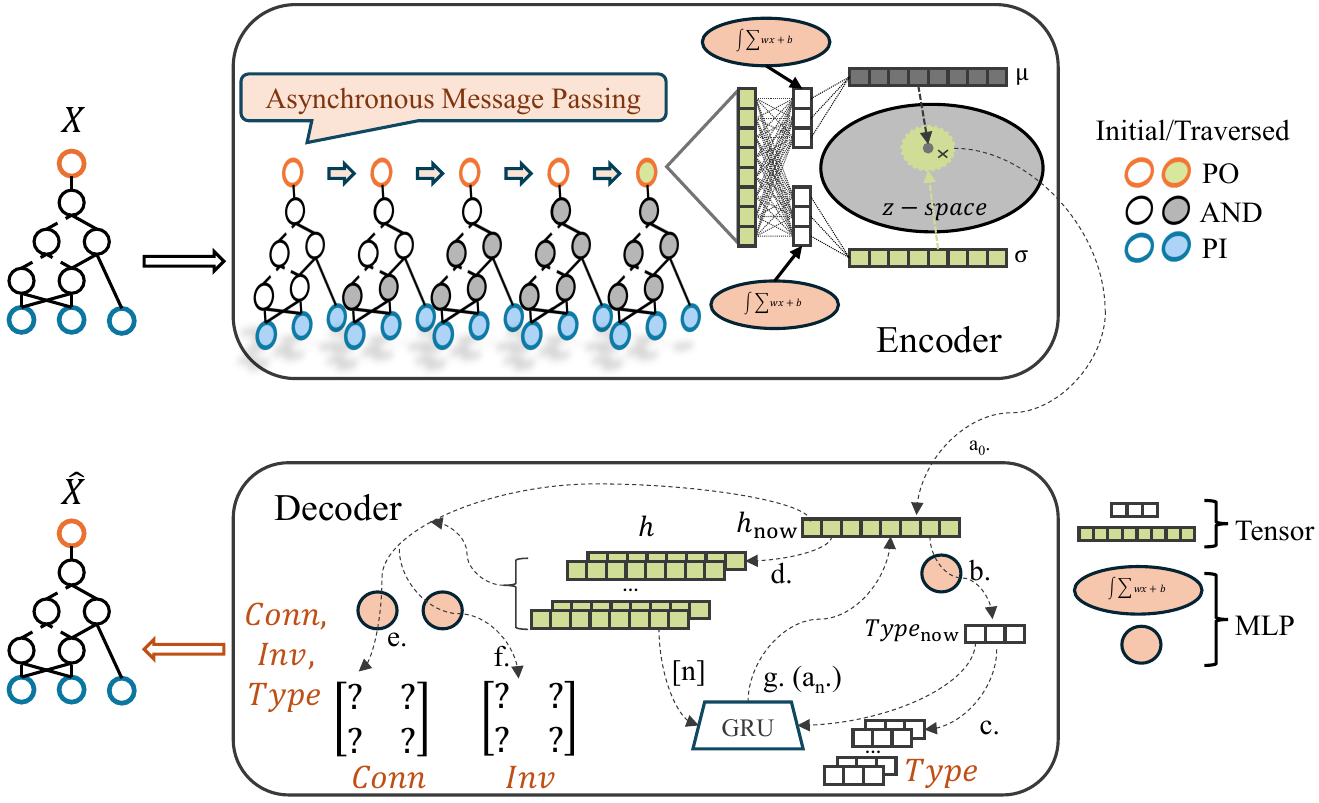}
    \caption{AIG-VAE Overview. %The figure illustrates the AIG-VAE framework with an Encoder (top) and Decoder (bottom). 
    The encoder uses asynchronous message passing to encode the input graph $X$ into a latent representation in $z$-space. The decoder reconstructs the graph $\hat{X}$ as follows:  
\textbf{($a_0$)} Sample a latent vector and initialize the first hidden state via an MLP;
\textbf{($b$)} Predict the node type (PI, PO, or AND) and assign PI to the first node;
\textbf{($c$)} Add the node type to the global tensor $Type$; 
\textbf{($d$)} Add the hidden state to the global tensor $h$;  
\textbf{($e$)} Predict connection edges with an MLP and store them in $Conn$; 
\textbf{($f$)} Predict inverters with an MLP and store them in $Inv$
\textbf{($g$, $a_n$)}; Update the hidden state for the next node using a GRU and repeat until all nodes are processed.  
The final AIG is reconstructed from $Conn$, $Inv$, and $Type$.}
    \label{AIG-VAE}

    \vspace{-2ex}

\end{figure}

Building on this foundation, we propose our machine learning model, the And-Inverter Graph Variational Autoencoder (AIG-VAE), specifically designed for encoding and decoding circuits represented as AIGs. As shown in \autoref{AIG-VAE}, similar to \autoref{subsec:VAE}, the model consists of two core components: Encoder and Decoder. %Their details are found below.

\subsubsection{Encoder} The encoder processes an input AIG \( X \) and maps it to a latent representation \( \mathbb{Z}_X = E(X) \), capturing both structural and functional features of the circuit. Our encoder employs an asynchronous message-passing mechanism across the AIG, allowing messages to propagate throughout the graph while preserving circuit hierarchy and dependencies. This model is designed for AIGs with a single primary output (PO), multiple primary inputs (PIs), and various AND gates. To initialize the process, each PI is encoded manually using one-hot encoding. The message-passing follows a breadth-first search (BFS) order, meaning each node does not receive or update its message until all its precedent nodes have been traversed. The message-passing and update process for each node \( v \) can be formalized as follows:
\begin{itemize}[wide, labelwidth=!, labelindent=0pt]
    \item \textbf{Message Aggregation}: For each AND node, we aggregate messages from its two preceding nodes \( u_1 \) and \( u_2 \). If an edge between a node and a predecessor includes an inverter, the incoming embedding from that predecessor is negated. The aggregated message \( m_v \) for node \( v \) is given by:
    \(
        m_v = \sum_{i=1}^{2} \text{sign}(e_i) \cdot h_{u_i}
    \label{eq:message_aggregation}
    \),
    where \( h_{u_i} \) represents the embedding of the \( i \)-th preceding node, and \( \text{sign}(e_i) \) is \( -1 \) if there is an inverter on the edge, \( +1 \) otherwise.

    \item \textbf{Node Update with GRU}: The aggregated message \( m_v \), combined with the node type \( t_v \) (indicating whether it’s a PI, PO, or AND), is fed into a Gated Recurrent Unit (GRU) \cite{cho2014learningphraserepresentationsusing} to update the node embedding. The GRU is a recurrent network cell designed to capture sequential and structural dependencies through gating mechanisms. Specifically, the GRU computes the next node embedding \( h_v \) in the following steps:
    \begin{enumerate}[wide, labelwidth=!, labelindent=0pt]
        \item \textbf{Update Gate (\( z_v \))}: Determines how much of the previous embedding to retain in the updated embedding:
        \(
        z_v = \text{sigmoid}(W_z m_v + U_z t_v + b_z)
        \).

        \item \textbf{Reset Gate (\( r_v \))}: Controls how much of the previous embedding to forget:
        \(
        r_v = \text{sigmoid}(W_r m_v + U_r t_v + b_r)
        \).

        \item \textbf{Candidate Embedding (\( \tilde{h}_v \))}: Computes the potential updated embedding:
        \(
        \tilde{h}_v = \tanh(W_h (r_v \odot h_{v-1}) + U_h t_v + b_h)
        \).

        \item \textbf{Final Node Embedding (\( h_v \))}: Combines the previous embedding \( h_{v-1} \) and the candidate embedding \( \tilde{h}_v \), weighted by the update gate \( z_v \):
        \(
        h_v = (1 - z_v) \odot h_{v-1} + z_v \odot \tilde{h}_v
        \).
    \end{enumerate}

        The notation used includes the sigmoid activation function, defined as \( \text{sigmoid}(x) = \frac{1}{1 + e^{-x}} \), and \( \tanh \), the hyperbolic tangent activation function defined as \(\tanh(x) = \frac{e^x - e^{-x}}{e^x + e^{-x}}\). The symbol \( \odot \) represents element-wise multiplication. The weight matrices \( W_z, W_r, W_h \) are for the aggregated message \( m_v \), while \( U_z, U_r, U_h \) represent the weight matrices for the node type \( t_v \). \( b_z, b_r, b_h \) denote bias terms.
        This allows the model to differentiate between node types while capturing both structural and functional dependencies of the graph. The node embedding \( h_v \) is computed as:
    \(
        h_v = \text{GRU}_e(m_v, t_v)
    \label{eq:node_update}
    \).
\end{itemize}

The embedding obtained from the final GRU update at the PO node is processed further to obtain the latent space representation. Specifically, this embedding, denoted as \( h_{\text{PO}} \), is passed through two separate Multi-Layer Perceptrons (MLPs) to produce the mean \( \mu \) and standard deviation \( \sigma \) for the VAE's latent distribution:
\(
    \mu = \text{MLP}_\mu(h_{\text{PO}})
    \label{eq:mu}
\), 
\(
    \sigma = \text{MLP}_\sigma(h_{\text{PO}})
    \label{eq:sigma}
\).
These parameters \( \mu \) and \( \sigma \) define the Gaussian distribution \( q(z|X) = \mathcal{N}(\mu, \sigma^2) \) from which the latent variable \( z \) is sampled. This approach enables the encoder to capture a probabilistic latent representation.

\subsubsection{Decoder}
The decoder reconstructs an AIG from the latent code \( \mathbb{Z}_X \), generating an output \( \hat{X} = D(\mathbb{Z}_X) \)  that preserves the essential characteristics of the original circuit while allowing for controlled variation\footnote{Controlled variation refers to the ability to introduce slight variations during decoding by sampling nearby latent codes within the distribution in latent space, enabling exploration of similar but not identical representations.}. A sample is drawn from the latent space using the mean \( \mu \) and standard deviation \( \sigma \) derived during encoding. During training, \( \sigma \) is used to augment data by adding stochasticity. However, during evaluation, only the mean \( \mu \) is used for consistency:
\(
    z \sim \mathcal{N}(\mu, \sigma^2)
\)

To initialize the node embeddings, \( z \) is passed through a fully connected linear layer followed by a tanh activation to produce the first node embedding \( h_1 \):
\(
    h_1 = \tanh(\text{Linear}(\mathbb{Z}_X))
\). The decoding process then unfolds as follows:
\begin{itemize}[wide, labelwidth=!, labelindent=0pt]
    \item \textbf{Node Type Prediction}: Using a MLP called \texttt{addNode}, the type of the node is determined from \( h \). This MLP classifies the node as either a Primary Input (PI), Primary Output (PO), or an AND gate:
    \(
        {Type}_i = \text{MLP}_{\text{addNode}}(h_i)
    \)

    \item \textbf{Connection and Inverter Determination}: For each node, two additional MLPs predict the connections to all previous nodes and whether there are inverters on these connections:
    \(
        {Connection}_{i,j} = \text{MLP}_{\text{connect}}(h_i, h_j)
    \) and  
    \(
        {Inverter}_{i,j} = \text{MLP}_{\text{inv}}(h_i, h_j)
    \),
    where \( h_i \) (\( h_j \)) represent the embeddings of current (previous) nodes.

    \item \textbf{Next Node Embedding}: To determine the embedding of the next node, the decoder uses a GRU specifically for decoding, denoted as \( \text{GRU}_d \). The GRU takes the current embedding \( h \) and the node type as inputs, producing the embedding for the next node:
    \(
        h_\text{i+1} = \text{GRU}_d(h_i, {Type}_i)
    \).
\end{itemize}
This process continues iteratively, building the entire AIG by adding nodes and determining connections, until the number of nodes matches that of the input circuit $X$ from which the latent space code encoded from.

\subsubsection{Loss Function} \label{subsubsec:loss}
When calculating the loss function, the AIG structure is represented by three primary tensors that capture essential circuit information: \(Type\) (which defines each node's type, such as PI, PO, or AND gate), \(Connection\) (an adjacency matrix indicating the presence of connections between nodes), and \(Inverter\) (indicating if an inverter is present on each edge). These tensors are calculated for the original circuit \( X \) and the reconstructed circuit \( \hat{X} \), denoted as \( Type_X \) and \( Type_{\hat{X}} \), \( Connection_X \) and \( Connection_{\hat{X}} \), and \( Inverter_X \) and \( Inverter_{\hat{X}} \), respectively.

For each tensor, we use the mean squared error (MSE) to compute the discrepancy between the original and reconstructed values. Here, \( N \) denotes the total number of nodes in the graph:
\(\mathcal{L}_{\text{Type}} = \frac{1}{N \cdot 3} \Sigma_{i=1}^{N} \Sigma_{j=1}^{3} ({Type}_{X, i, j} - {Type}_{\hat{X}, i, j})^2,\)
\(\mathcal{L}_{\text{Connection}} = \frac{1}{N^2} \Sigma_{i=1}^{N} \Sigma_{j=1}^{N} ({Connection}_{X, i, j} - {Connection}_{\hat{X}, i, j})^2,\)
\(\mathcal{L}_{\text{Inverter}} = \frac{1}{N^2} \Sigma_{i=1}^{N} \Sigma_{j=1}^{N} ({Inverter}_{X, i, j} - {Inverter}_{\hat{X}, i, j})^{2}\).

% For our implementation, we set \( \alpha = 0.3 \), \( \beta = 0.3 \), \( \gamma = 0.3 \), and \( \delta = 0.1 \) to balance the contributions of each component in the loss function. 

The KL divergence loss \( \mathcal{L}_{{KL}} \) regularizes the latent space by measuring the divergence between the encoded distribution \( q(z|X) = \mathcal{N}(\mu, \sigma^2) \) and a prior Gaussian distribution \( p(z) = \mathcal{N}(0, I) \). It is computed as:
\(
\mathcal{L}_{{KL}} = D_{{KL}}(q(z|X) \parallel p(z)) = \frac{1}{2} \sum_{i=1}^d \left( \sigma_i^2 + \mu_i^2 - \log(\sigma_i^2) - 1 \right)
\),
where \( d \) is the dimensionality of the latent space, and \( \mu \) and \( \sigma \) are the mean and standard deviation vectors from the encoder.

The overall loss function \( \mathcal{L} \) is then defined as a weighted sum of these individual components, along with the KL divergence loss (\( \mathcal{L}_{\text{KL}} \)) to regularize the latent space:
\(
\mathcal{L} = \alpha \cdot \mathcal{L}_{\text{Type}} + \beta \cdot \mathcal{L}_{\text{Connection}} + \gamma \cdot \mathcal{L}_{\text{Inverter}} + \delta \cdot \mathcal{L}_{\text{KL}}.
\)
Here, \( \alpha, \beta, \gamma \), and \( \delta \) are hyperparameters that control the relative contribution of each loss component to the overall objective. This loss function ensures that the reconstructed circuit retains the original structure and functionality while adhering to the latent space's constraints. For our implementation, we set \( \alpha = 0.3 \), \( \beta = 0.3 \), \( \gamma = 0.3 \), and \( \delta = 0.1 \) to balance the contributions of each component in the loss function.

\subsection{Interpolation between Function and Appearance}\label{subsec:interpolation}
    In \autoref{subsec:MethodOverview}, we briefly introduced the concept of interpolating between the latent space representations of a functional circuit and the appearance circuit, enabling us to examine the intermediate states between the two. In this section, we will discuss this in greater detail.
    
\begin{figure}[t]
    \centering
    \includegraphics[width=0.98\linewidth]{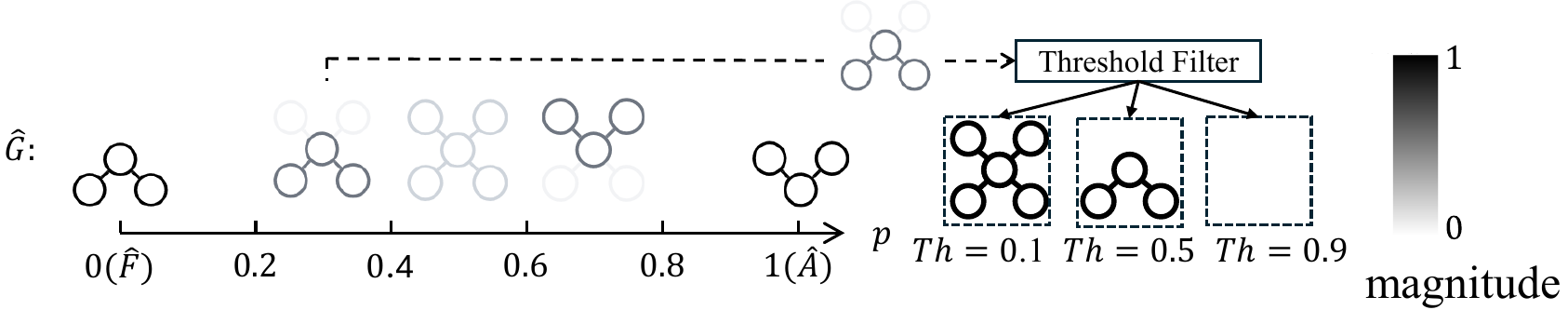}
    \caption{Interpolation between $\hat{F}$ and $\hat{A}$ and impact of threshold values on binary decision-making. The intensity of the color indicates the magnitude.}
    \label{Interpolation}

    \vspace{-2ex}
    
\end{figure}

    The interpolation process blends the functional circuit \( F \) and appearance circuit \( A \) by leveraging their respective latent space representations, encoded as \( \mathbb{Z}_F \) and \( \mathbb{Z}_A \). These latent representations are interpolated based on a proportion factor \( p \), defined :
\(
\mathbb{Z}_G = (1 - p) \mathbb{Z}_F + p \mathbb{Z}_A
\), 
where \( \mathbb{Z}_G \) balances between the latent representations of the functional and appearance circuits. The decoder then reconstructs a circuit \( \hat{G} \) from \( \mathbb{Z}_G \). Here, \( p \in [0, 1] \) determines the interpolation factor. When \( p = 0 \), \( \mathbb{Z}_G \) focuses entirely on the characteristics of the functional circuit, retaining the function and appearance of \( F \). When \( p = 1 \), \( \mathbb{Z}_G \) is dominated by the characteristics of the appearance circuit. For intermediate values of \( p \), the generated circuit shows a blend of function and appearance of functional and appearance circuits, as illustrated in \autoref{Interpolation}. 

Once the interpolated latent vector \( \mathbb{Z}_G \) is obtained, it is decoded using the decoder to produce the new circuit 
\(
\hat{G} = D(Z_G)
\label{eq:decode}
\).
Due to the interpolation process, the rebuilt \( \hat{G} \) varies as different proportion values are selected. In \autoref{Interpolation}, the lighter color in the figure indicates that the blended values decrease in magnitude. These decreased values do not directly indicate clear node connections or inverters. Therefore, a threshold value \( Th \) is applied to convert these values into binary decisions using a threshold filter, as shown in \autoref{filter}. This filtering process ensures that the generated circuit has clear, interpretable connections and maintains structural integrity, even when blending between functional circuit and appearance circuit.

The final circuit, after applying thresholding, achieves the desired balance between functionality and appearance. By adjusting the interpolation factor \( p \) and the threshold \( Th \), designers can fine-tune the generated circuit to better meet specific requirements.

    \subsection{Functional Preserve and Appearance Mimicking}\label{subsec:FPandAM}
    The generated AIG \( \hat{G} \), derived from the latent space distribution between the functional circuit \( F \) and the appearance circuit \( A \), exhibits differences in both functionality and appearance compared to \( F \) and \( A \). To reconcile these differences, we employ two fixes: \textit{Functional Preserve}, which restores the functional behavior of \( F \) in \( \hat{G} \), and \textit{Appearance Mimicking}, which further adjusts \( \hat{G} \) to closely resemble \( A \).

        \subsubsection{New Covert Gates} \label{FBFIUT}

To implement these fixes, we introduce specialized components designed using the covert gate methodology. These camouflaged components leverage a unique manufacturing technology that enables transistors to remain in an always-on or always-off state \cite{shakya_covert_2019}. This capability allows the creation of gates that deceive reverse engineers by emitting misleading SEM images while preserving the circuit's true functionality. 

\begin{figure}[t]
    \centering\includegraphics[width=1\linewidth]{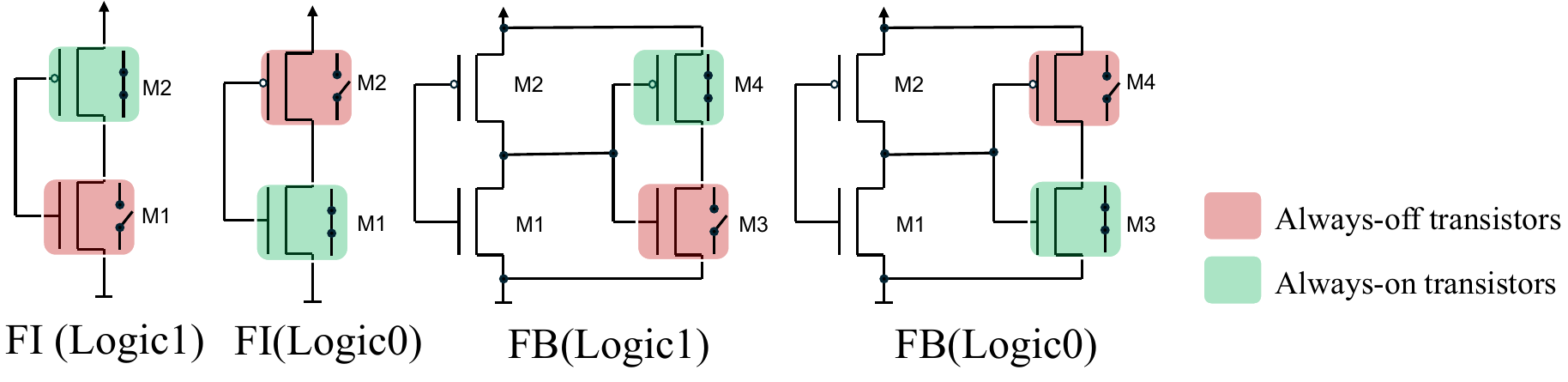}
    \caption{CMOS circuits of Fake Inverter (FI) and Fake Buffer (FB). }
    \label{FIandFB}
    \vspace{-3ex}
\end{figure}

\begin{figure}[t]
    \centering \includegraphics[width=0.75\linewidth]{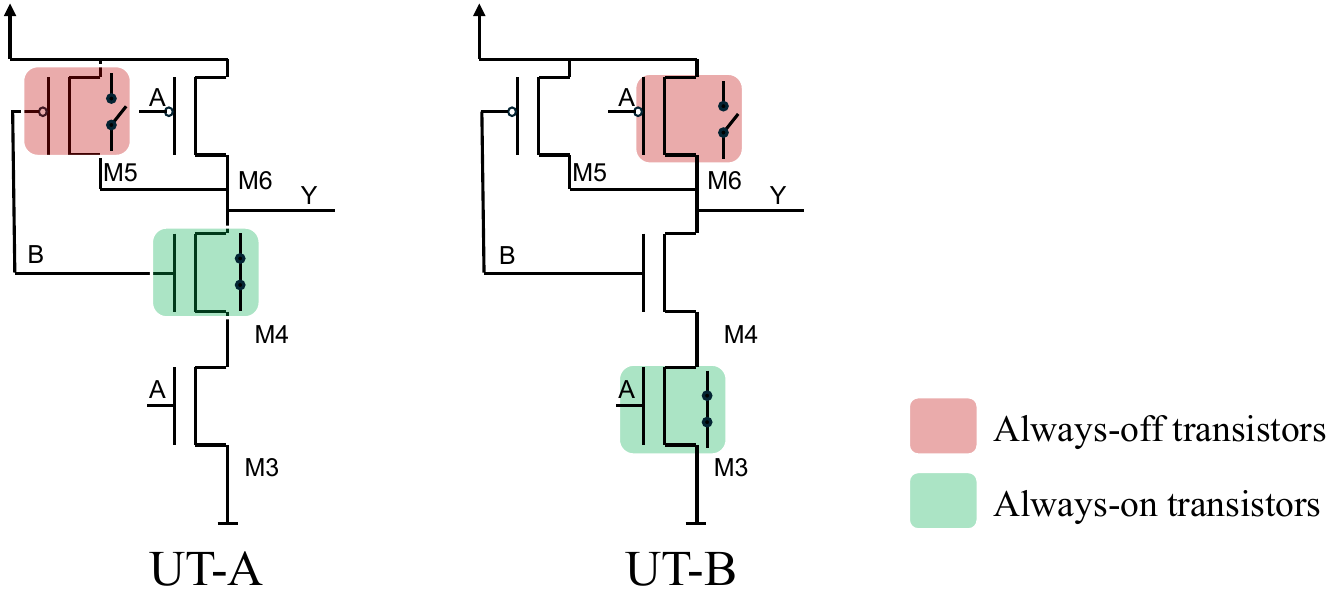}
    \caption{CMOS circuits of Universal Transmitter (UT). From left to right: UT-A, UT-B.}
    \label{UT}
\vspace{-3ex}    
\end{figure}

We have developed three types of components based on covert gates: the \textbf{Fake Inverter (FI)}, \textbf{Fake Buffer (FB)}, and \textbf{Universal Transmitter (UT)}. These components would be strategically incorporated during the Functional Preserve and Appearance Mimicking stages to obscure the circuit's true logic while presenting a deceptive external appearance.
As shown in \autoref{FIandFB}, fake components, the FI and FB, are functionally distinct from the corresponding real component, but they maintain the same external appearance. The Universal Transmitter (UT), depicted in \autoref{UT}, has two variations (UT-A and UT-B), having contrasting functions but the same appearance.
\begin{table}[t]
    \centering
    \caption{Function and Appearance of FI, FB, and UT}
    \label{FI_FB_UT_Table}
    \scriptsize
    \begin{tabular}{ccc}
    \toprule
    \textbf{Component} & \textbf{Function} & \textbf{Appearance} \\
    \midrule
    FI     & logic1 / logic0 & 1 Inverter \\
    FB     & logic1 / logic0 & 2 Inverters \\
    UT-A   & Buffer / logic 1 / logic 0           & 1 NAND \\
    UT-B   & Inverter / logic 1/ logic 0         & 1 NAND \\
    \bottomrule
    \end{tabular}
\end{table}
As shown in \autoref{FI_FB_UT_Table}, the various types of Fake Inverters (FI), Fake Buffers (FB), and Universal Transmitters (UT) provide distinct functionalities while maintaining consistent external appearances. These components are specifically designed to camouflage individual signals, effectively concealing their true functions from their outward appearance. For FI and FB, a signal can be made to appear inverted or non-inverted while functionally being logic1/logic0. For UT, the two inputs can be selected like in a MUX, while maintaining the appearance of a NAND gate. When used in the fix step, UT can choose between a signal and its negation, effectively functioning as a buffer or inverter -- both sharing the same appearance. Like FI and FB, UT can also be configured to output constant logic1/logic0.
% the functionality of a buffer and an inverter can be combined into a single, visually identical component, further complicating RE efforts.

\subsubsection{Functional Preservation and Appearance Mimicking} \label{FPAME}
Building on the introduction of these covert gates -- Fake Inverters (FI), Fake Buffers (FB), and Universal Transmitters (UT) -- this section explains their integration into the processes of Functional Preservation and Appearance Mimicking.

The Functional Preservation process ensures that the generated circuit \( \hat{G} \) functions identically to the original functional circuit \( F \), while effectively camouflaging its true purpose. This is achieved by comparing the structure and behavior of \( \hat{G} \) to \( F \) and making the necessary corrections. The process begins by equalizing the size of \( \hat{G} \) and \( F \) by adding dummy nodes, ensuring that both circuits have the same number of gates and nodes across all three types of nodes (PI, PO and AND). Subsequently, the connections and logic of each node in \( \hat{G} \) are compared with those in \( F \), and any discrepancies in logic or connections are identified and corrected. To address these discrepancies, Fake Inverter (FI), Fake Buffer (FB), and Universal Transmitter (UT) components are applied. These components enable the adjustments required to make \( \hat{G} \) functionally equivalent to \( F \) while maintaining its camouflaged appearance.

With each corrective step applied to a signal in \( \hat{G} \), \autoref{FixStepTable} compares the generated circuit \( \hat{G} \) with the functional circuit \( F \), detailing the fix operations and their effects. The values \( 00/01 \), \( 10 \), and \( 11 \) represent the states of connections and inversions: the first bit indicates the presence of a connection (0 for none, 1 for a connection), and the second bit indicates inversion (0 for no inversion, 1 for inversion). The ``Fix Step'' column outlines the actions taken, such as adding connections or applying FI, FB, or UT components. ``Connect" and ``Insert INV" means we have to restore the function with no covert gate applicable in this situation.  ``N/A'' indicates that no fix is necessary.

\begin{table}[t]
    \centering
    \caption{Fix Steps for Functional Preservation and Appearance Mimicking}
    \label{FixStepTable}\scriptsize
    \begin{tabular}{ccc|ccc}
        \toprule
        \multicolumn{3}{c|}{\textbf{Functionality Preservation}} & \multicolumn{3}{c}{\textbf{Appearance Mimicking}} \\
        \textbf{\( \hat{G} \)} & \textbf{Reference} \(F\) & \textbf{Fix Step} & \textbf{\( \hat{G}_F \)} & \textbf{Reference} \(A\) & \textbf{Fix Step} \\
        \midrule
        00/01 & 00/01 & N/A         & 00/01 & 00/01 & N/A \\
        00/01 & 10    & Connect     & 00/01 & 10    & FB \\
        00/01 & 11    & Insert INV  & 00/01 & 11    & FI \\
        10    & 00/01 & FB          & 10    & 00/01 & N/A \\
        10    & 10    & N/A         & 10    & 10    & N/A \\
        10    & 11    & UT-B        & 10    & 11    & UT-A \\
        11    & 00/01 & FI          & 11    & 00/01 & N/A \\
        11    & 10    & UT-A        & 11    & 10    & UT-B \\
        11    & 11    & N/A         & 11    & 11    & N/A \\
        \bottomrule
    \end{tabular}

    \vspace{-3ex}
    
\end{table}

Following the Functional Preservation is the Appearance Mimicking. \( \hat{G}_F \), the output of Functional Preservation, and \( A \) are padded to the same size. Their edges are then compared, and the necessary fix steps are applied, treating \( A \) as the desired appearance and \( \hat{G}_F \) as the desired function. The final circuit \( \hat{G}_{F,A} \) is produced after Appearance Mimicking, which retains the functionality of the functional circuit while presenting an appearance similar to the appearance circuit.

\section{Experimental Results} \label{sec:results}

This section provides a comprehensive evaluation of IP Camouflage. 
% Through a series of experiments, we assessed the ability of our AIG-VAE model to encode and reconstruct circuits accurately. We further analyze the effectiveness of the framework in terms of its resilience against SAT-based attacks and its ability to deceive adversaries performing RE. 
Note that all the circuits produced by IP Camouflage successfully pass formal verification with the original circuit design.
The experiments address three objectives:
\begin{enumerate}[wide,  labelindent=0pt]
    \item \textbf{Validate Functional Understanding of the Model}: In \autoref{subsec::EvalonModel}, we evaluate how well the AIG-VAE model captures the functional characteristics of circuits by correlating latent space distances with structural differences.
    \item \textbf{SAT-Attack Resilience Analysis}: In \autoref{subsec::EvalSAT}, we analyze the effectiveness of IP camouflage applying covert gate components against SAT-based attacks, which are widely acknowledged as a powerful and practical threat model in hardware security.
    \item \textbf{Demonstrated Robustness Against AI-Enhanced Reverse Engineering Tools}: In \autoref{subsec:GNN_RE}, we evaluate the robustness of our camouflaged circuits against advanced RE techniques by applying Graph Neural Networks (GNNs)~\cite{alrahis_gnn_re_2022} to classify circuit nodes.

\end{enumerate}

    \subsection{AIG-VAE Model Training and Evaluation}\label{subsec::EvalonModel}

Our dataset comprises combinational circuits from the ISCAS85 \cite{ISCAS} and EPFL \cite{EPFL} benchmarks, converted into AIGs for uniform input representation suitable for the AIG-VAE model. Note that the former benchmarks were originally utilized for an RE case study. Output signals and their preceding signals are decomposed into tree graphs, which serve as model inputs. To manage complexity, trees exceeding a maximum node limit are excluded. The dataset is split into training and testing subsets (80-20 split). \autoref{tab:dataset_summary} summarizes the dataset with details on the number of graphs and nodes.
Graph processing is facilitated by the Deep Graph Library (DGL) \cite{DGL}, which integrates seamlessly with PyTorch and handles graphs as directed acyclic graphs (DAGs).

\begin{table}[t]
    \centering
        \scriptsize
    \caption{Dataset Summary for Training and Testing}
    \label{tab:dataset_summary}
    \begin{tabular}{lcc}
        \toprule
        \textbf{Dataset Subset} & \textbf{Number of Graphs} & \textbf{Total Number of Nodes} \\ 
        \midrule
        Training Dataset & 1,482 & 63,091 \\ 
        Testing Dataset  & 371   & 14,515 \\ 
        \midrule
        \textbf{Total}   & 1,853 & 77,606 \\ 
        \bottomrule
    \end{tabular}

    \vspace{-3ex}
    
\end{table}

The AIG-VAE model, implemented in PyTorch (version 2.3.1) \cite{Pytorch}, was trained on an NVIDIA A100 GPU with 80 GB memory. Using the Adam optimizer with a learning rate of 0.0001 and batch size of 1, training aimed to minimize a combined loss function (reconstruction loss, structural consistency, and latent space regularization; see \autoref{subsubsec:loss}). The model was trained for 100 epochs with early stopping based on validation loss.
Key training hyperparameters include a latent space dimension of 512, a KL divergence weight (\( \delta \)) of 0.1, and reconstruction loss weights (\( \alpha / \beta / \gamma \)) of 0.3 each.

We evaluate the AIG-VAE model’s ability to encode and capture the structural and functional characteristics of circuits by analyzing the relationship between \textit{graph edit distance (GED)} and \textit{latent space distance (LSD)} for all pairwise combinations of graphs in the testing dataset. GED, computed using the \texttt{Networkx} library \cite{networkx}, measures structural similarity as the minimum number of edit operations (node insertion, deletion, or substitution) required to transform one graph into another. LSD is the Euclidean distance between the latent representations of two graphs produced by the AIG-VAE encoder.

\begin{figure}[t]
    \centering
    \includegraphics[width=0.98\linewidth]{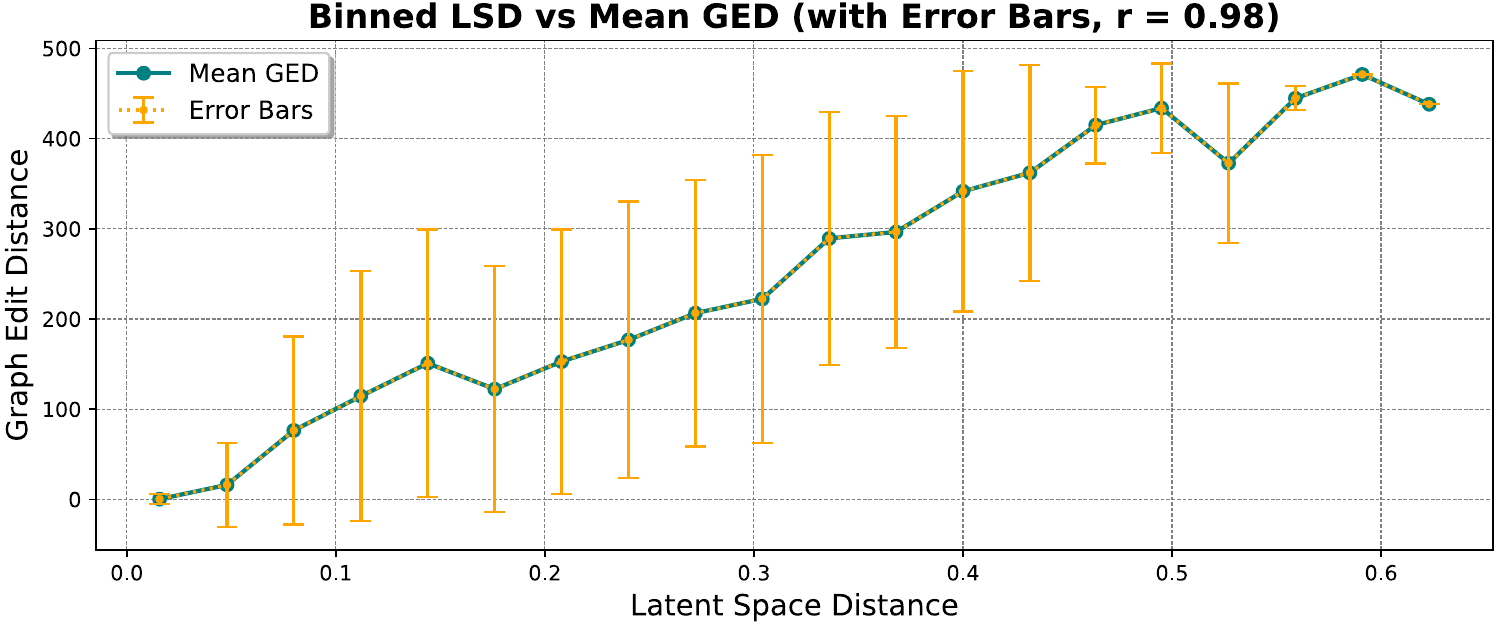}
    \caption{Binned plot of latent space distance (LSD) vs. mean graph edit distance (GED) for all pairwise combinations of graphs in the testing dataset.}
    \label{fig:latent_vs_ged}

    \vspace{-1ex}
    
\end{figure}

Given the computational cost of GED, a timeout was applied, and data points exceeding the limit were discarded. Following this, we gathered 136,445 valid data points. To better interpret the relationship between GED and LSD, we use a \textit{binned analysis}, dividing LSDs into 20 bins and computing the mean GED and standard deviation for each bin. The results, shown in \autoref{fig:latent_vs_ged}, reveal a strong positive correlation: larger LSDs correspond to greater GED values, as indicated by the trend and a Pearson correlation coefficient of \( r = 0.98 \), with some errors.
\textit{This analysis demonstrates that the AIG-VAE effectively encodes features into the latent space, capturing both structural differences and functional characteristics.} These latent representations provide a meaningful and continuous basis for downstream tasks.

\subsection{Evaluating Resistance to SAT-Guided Reverse Engineering} \label{subsec::EvalSAT}

SAT-based attacks pose a major threat to logic locking (LL) and camouflaging by exploiting oracle access and iterative SAT solving to recover circuit functionality.
To evaluate SAT resilience, we model covert gates -- Fake Inverter (FI), Fake Buffer (FB), and Universal Transmitter (UT) -- as key-programmable elements, each controlled by two keys to select logic-1, logic-0, or normal behavior. Gates with the same appearance as covert gates are also treated as potential targets, greatly expanding the key space.

We tested four circuit pairs (Table~\ref{tab:chosenAF}), each combining a functional and an appearance circuit. For each, we generated multiple IP Camouflage configurations with varying thresholds and compared them to logic-locked versions with matched area overhead. As shown in Table~\ref{tab:sat_vs_logiclocking}, IP Camouflage \textit{consistently produces significantly more keys under similar area, power, and delay overheads} -- enhancing resistance due to the exponential nature of SAT solving.

To assess solver run time, we ran SAT attacks on configurations with maximum and minimum key counts per pair. As reported in Table~\ref{tab:sat_vs_logiclocking}, IP Camouflage circuits often caused timeouts or memory exhaustion on a 100~GB RAM, AMD EPYC 7702 64-core server using the Glucose SAT solver, while LL equivalents were typically solved within milliseconds. These results confirm the practical strength of IP Camouflage against SAT-guided RE.

\begin{table}[t]
    \centering
    \caption{Circuit Pairs Used for SAT-Guided Reverse Engineering Evaluation}
    \scriptsize
    \begin{tabular}{lcccc}
        \toprule
        \textbf{Pair} & \textbf{Circuit \( A \)} & \textbf{\#\( Nodes \)} & \textbf{Circuit \( F \)} & \textbf{\#\( Nodes \)} \\ 
        \midrule
        0 & c5315\_N7705 & 132 & c432\_N421 & 166 \\ 
        1 & c5315\_N7504 & 144 & i2c\_po061 & 53 \\ 
        2 & banyan\_8\_out\_5 & 125 & c3540\_N4589 & 122 \\ 
        3 & c2670\_N3809 & 154 & memctrl\_po0198 & 149 \\ 
        \bottomrule
    \end{tabular}
    \label{tab:chosenAF}

    \vspace{-3ex}
    
\end{table}

\begin{table*}[!t]
    \centering
\caption{
IP Camouflage vs. Logic Locking for number of keys and overhead across Pairs 0–3.
\textbf{Note:} Only edge-case configurations were considered.
\textbf{P}: Pair index.
\textbf{Th}: Threshold parameter for IP Camouflage.
\textbf{A}, \textbf{Pwr}, \textbf{Dly}:Area, Power, and Delay Overhead from IP Camouflage.
\textbf{\#K}: Key count from IP Camouflage.
\textbf{\#K\textsubscript{eq} / A}: Number of keys required for logic locking to match the area overhead.
\textbf{T\textsubscript{Ours}}: SAT Attack Time of IP Camouflage.
\textbf{T\textsubscript{LL}}: SAT Attack Time of Logic Locking.
\textbf{\textcolor{red}{TO}}: Time out for 24-hour limit.
\textbf{\textcolor{red}{OoM}}: Out of memory for 100 GB limit.
}

    \label{tab:sat_vs_logiclocking}
    \scriptsize
    \begin{tabular}{c c c c c c c c c| c c c c c c c c c}
        \toprule
        \textbf{P} & \textbf{Th} & \textbf{A} & \textbf{Pwr} & \textbf{Dly} & \textbf{\#K} & \textbf{\#K\textsubscript{eq} / AO} & \textbf{T\textsubscript{Ours}} & \textbf{T\textsubscript{LL}}
 &
        \textbf{P} & \textbf{Th} & \textbf{A} & \textbf{Pwr} & \textbf{Dly} & \textbf{\#K} & \textbf{\#K\textsubscript{eq} / AO} & \textbf{T\textsubscript{Ours}} & \textbf{T\textsubscript{LL}}
 \\
        \midrule
        0 & 0.01 & 1.40× & 1.37× & 1.00× & 332 & 68 / 1.41× & \textcolor{red}{TO} & 670 ms & 2 & 0.01 & 1.34× & 1.33× & 1.00× & 244 & 44 / 1.34× & \textcolor{red}{TO} &  89 ms\\
        0 & 0.02 & 1.38× & 1.34× & 1.00× & 318 & 64 / 1.39× & - & -  & 2 & 0.02 & 1.32× & 1.31× & 1.00× & 238 & 44 / 1.34× & - &  -\\
        0 & 0.03 & 1.37× & 1.32× & 1.00× & 314 & 60 / 1.36× & - & - & 2 & 0.03 & 1.29× & 1.29× & 1.00× & 232 & 36 / 1.28× & - &  -\\
        0 & 0.04 & 1.34× & 1.25× & 1.00× & 288 & 56 / 1.34× & - & - & 2 & 0.04 & 1.24× & 1.24× & 1.00× & 216 & 32 / 1.24× & - &  -\\
        0 & 0.05 & 1.32× & 1.18× & 1.05× & 268 & 52 / 1.30× & - & - & 2 & 0.05 & 1.20× & 1.20× & 1.00× & 204 & 28 / 1.22× & - &  -\\
        0 & 0.06 & 1.26× & 1.12× & 1.21× & 248 & 44 / 1.27× & - & - & 2 & 0.06 & 1.17× & 1.17× & 1.00× & 196 & 24 / 1.18× & - &  -\\
        0 & 0.07 & 1.26× & 1.12× & 1.21× & 248 & 40 / 1.25× & - & - & 2 & 0.07 & 1.17× & 1.17× & 1.00× & 196 & 24 / 1.18× & - &  -\\
        0 & 0.08 & 1.12× & 1.05× & 1.21× & 218 & 20 / 1.12× & - & - & 2 & 0.08 & 1.16× & 1.16× & 1.00× & 192 & 20 / 1.15× & - &  -\\
        0 & 0.09 & 1.08× & 1.04× & 1.21× & 212 & 12 / 1.08× & 3525 s & 35 ms & 2 & 0.09 & 1.16× & 1.15× & 1.00× & 190 & 20 / 1.15× & 10370 s &  48 ms\\
        \midrule
        1 & 0.01 & 1.18× & 1.18× & 1.00× & 104 & 20 / 1.20× & 156 s & 14 ms & 3 & 0.01 & 1.30× & 1.30× & 1.00× & 262 & 44 / 1.30× & \textcolor{red}{OoM} &  73 ms\\
        1 & 0.02 & 1.15× & 1.15× & 1.00× & 98  & 16 / 1.16× & - & - & 3 & 0.02 & 1.28× & 1.29× & 1.00× & 256 & 40 / 1.27× & - &  -\\
        1 & 0.03 & 1.14× & 1.14× & 1.00× & 94  & 16 / 1.16× & - & - & 3 & 0.03 & 1.26× & 1.26× & 1.00× & 248 & 40 / 1.27× & - &  -\\
        1 & 0.04 & 1.14× & 1.13× & 1.00× & 92  & 16 / 1.16× & - & - & 3 & 0.04 & 1.24× & 1.25× & 1.00× & 242 & 36 / 1.25× & - &  -\\
        1 & 0.05 & 1.12× & 1.11× & 1.00× & 88  & 12 / 1.12× & - & - & 3 & 0.05 & 1.21× & 1.21× & 1.00× & 230 & 32 / 1.22× & - &  -\\
        1 & 0.06 & 1.12× & 1.10× & 1.00× & 86  & 12 / 1.12× & - & - & 3 & 0.06 & 1.21× & 1.21× & 1.00× & 228 & 28 / 1.19× & - &  -\\
        1 & 0.07 & 1.12× & 1.10× & 1.00× & 86  & 12 / 1.12× & - & - & 3 & 0.07 & 1.21× & 1.20× & 1.00× & 226 & 32 / 1.22× & - &  -\\
        1 & 0.08 & 1.11× & 1.10× & 1.00× & 84  & 12 / 1.12× & - & - & 3 & 0.08 & 1.16× & 1.14× & 1.00× & 204 & 24 / 1.16× & - &  -\\
        1 & 0.09 & 1.10× & 1.09× & 1.00× & 82  & 12 / 1.12× & 247 s & 14 ms & 3 & 0.09 & 1.08× & 1.08× & 1.00× & 182 & 12 / 1.08× & \textcolor{red}{OoM} &  16 ms\\
        \bottomrule
    \end{tabular}

\vspace{-2ex}
    
\end{table*}

% \begin{table}[ht]
%     \centering
%     \caption{SAT Attack Time Comparison of IP Camouflage and LL}
%     \label{tab:sat_vs_logiclocking}
%     \scriptsize
%     \begin{tabular}{c|c|c|c|c}
%         \toprule
%         \textbf{Pair} & \textbf{\#Keys (Ours)} & \textbf{\#Keys (LL)} & \textbf{SAT Time (Ours)} & \textbf{SAT Time (LL)} \\
%         \midrule
%         0 & 332 & 68  & TIME OUT & 670 ms\\
%         0 & 212 & 12  & 3525 s & 35 ms\\
%         1 & 104 & 20  & 156 s & 14 ms\\
%         1 & 82  & 12  & 247 s & 14 ms\\
%         2 & 244 & 44  & TIME OUT & 89 ms\\
%         2 & 190 & 20 & 10370 s & 48 ms \\
%         3 & 262 & 44 & Out of Memory & 73 ms \\
%         3 & 182 & 12 & Out of Memory & 16 ms \\
%         \bottomrule
%     \end{tabular}
% \end{table}

\subsection{GNN-RE Analysis for Post-Camouflage Node Classification} \label{subsec:GNN_RE}

To evaluate the effectiveness of the proposed IP Camouflage methodology, we employed GNN-RE \cite{alrahis_gnn_re_2022}, which uses state-of-the-art node classification based on GraphSAINT \cite{zeng_graphsaint_2020}. The dataset includes eight benchmark circuits from ISCAS85 \cite{ISCAS} and EPFL \cite{EPFL}, with a fixed training set across all experiments as shown in \autoref{tab:DatasetOverview}.
\begin{table}[t]
    \centering
    \caption{GNN-RE Dataset Configuration. A shared training set is used for all groups. Test pairs indicate functional (\textcolor{softgreen}{green}) vs. appearance (\textcolor{softred}{red}) circuits.}
    \label{tab:DatasetOverview}
    \scriptsize
    \begin{tabular}{c|l}
        \toprule
        \textbf{Split} & \textbf{Circuits (Signal)} \\
        \midrule
        \textbf{Training} &
        \begin{tabular}[c]{@{}l@{}}
            banyan (8\_out\_5), memctrl (po0501), c2670 (N3809), c3540 (N4589), \\
            i2c (po061), c5315 (N7504), c7552 (N10351), c432 (N421)
        \end{tabular} \\
        \midrule
        \textbf{Group 1} &
        Val: memctrl (po0198), c5315 (N7705) \\
        & Test: \textcolor{softgreen}{memctrl}/\textcolor{softred}{c5315} (po0198 / N7504), \textcolor{softgreen}{c5315}/\textcolor{softred}{c432} (N7705 / N421) \\
        \midrule
        \textbf{Group 2} &
        Val: i2c (po059), c432 (N430) \\
        & Test: \textcolor{softgreen}{i2c}/\textcolor{softred}{c2670} (po059 / N3809), \textcolor{softgreen}{c432}/\textcolor{softred}{memctrl} (N430 / po0472) \\
        \midrule
        \textbf{Group 3} &
        Val: c7552 (N10110), c5315 (N7705) \\
        & Test: \textcolor{softgreen}{c7552}/\textcolor{softred}{i2c} (N10110 / po061), \textcolor{softgreen}{c5315}/\textcolor{softred}{memctrl} (N7705 / po0501) \\
        \midrule
        \textbf{Group 4} &
        Val: c7552 (N10110), c432 (N430) \\
        & Test: \textcolor{softgreen}{c7552}/\textcolor{softred}{memctrl} (N10110 / po0501), \textcolor{softgreen}{c432}/\textcolor{softred}{c5315} (N430 / N7504) \\
        \bottomrule
    \end{tabular}

\vspace{-3ex}
    
\end{table}
For each group, the test set includes camouflaged signals (functional) and their deceptive counterparts (appearance). The proportional parameter \(p\) was varied across \{0.1, 0.3, 0.5, 0.7, 0.9\}, and three F1-scores were computed:
\begin{itemize}[wide, labelindent=0pt]
    \item $F1_{\text{Valid}}$: Real-label classification accuracy on the validation set.
    \item \textcolor{softgreen}{$F1_{\text{Cryptic}}$}: Test classification accuracy for functional circuit label.
    \item \textcolor{softred}{$F1_{\text{Mimetic}}$}: Test classification accuracy for appearance circuit label.
\end{itemize}

Cryptic camouflage is achieved when \(\textcolor{softgreen}{F1_{\text{Cryptic}}} < F1_{\text{Valid}}\), indicating reduced recognition of functional nodes. Mimetic camouflage is considered successful when \(\textcolor{softred}{F1_{\text{Mimetic}}}\) significantly exceeds random guess and is high enough to be distinguishable. To further highlight the improvement, we compare our method to the random covert gate insertion strategy proposed in \cite{shakya_covert_2019}, using two baselines: (1) random insertion with area overhead matched to ours and (2) 5\% gate-level random insertion originally used in~\cite{shakya_covert_2019}. 

The results in \autoref{tab:GNN_RE_Random_Compare} demonstrate that our proposed IP Camouflage \textit{achieves both significantly lower \textcolor{softgreen}{\(F1_{\text{Cryptic}}\)} and higher \textcolor{softred}{\(F1_{\text{Mimetic}}\)} scores compared to the baseline of random covert gate insertion.} This dual effect indicates not only reduced classification accuracy of functional (true) circuits, but also increased misclassification as appearance (decoy) circuits. Such performance highlights the effectiveness of our structured, model-guided camouflage methodology in misleading GNN-based reverse engineering. Compared to the 5\% random insertion scheme proposed in \cite{shakya_covert_2019}, our approach provides a stronger functional concealment and adversarial deception, under similar or lower area overhead.

\begin{table}[t]
    \centering
    \renewcommand{\arraystretch}{0.75}
    \caption{
        GNN-RE F1-score comparison across four groups and five \(p\) values. 
        \textbf{$F1_\text{Val}$}: Validation accuracy.
        \textcolor{softgreen}{$F1_\text{Cryptic}$} (lower = better),
        \textcolor{softred}{$F1_\text{Mimetic}$} (higher = better).
        Superscripts: \textbf{Ours} = IP Camouflage, \textbf{Rand5\%} = 5\% CG insertion from \cite{shakya_covert_2019}, \textbf{RandAM} = area-matched random insertion. The F1-score for random guess is 0.125.
    }
    \label{tab:GNN_RE_Random_Compare}
    \scriptsize
    \begin{tabular}{c|c|ccccc}
        \toprule
        \textbf{Group} & \textbf{Metric} & \(p=.1\) & \(p=.3\) & \(p=.5\) & \(p=.7\) & \(p=.9\) \\
        \midrule
        {1}
        & $F1_\text{Val}$ & 0.59 & 0.58 & 0.59 & 0.58 & 0.58 \\
        \cmidrule(lr){1-7}
        {\textcolor{softgreen}{$F1_\text{Cryptic}$}} 
        & Ours    & \textbf{0.50} & 0.54 & \textbf{0.52} & \textbf{0.46} & \textbf{0.48} \\
        & Rand5\% & 0.58 & 0.58 & 0.58 & 0.58 & 0.58 \\
        & RandAM  & 0.52 & \textbf{0.53} & 0.53 & 0.52 & 0.53 \\
        \cmidrule(lr){1-7}
        {\textcolor{softred}{$F1_\text{Mimetic}$}} 
        & Ours    & \textbf{0.28} & \textbf{0.26} & \textbf{0.28} & \textbf{0.42} & \textbf{0.39} \\
        & Rand5\% & 0.12 & 0.12 & 0.12 & 0.12 & 0.12 \\
        & RandAM  & 0.18 & 0.14 & 0.18 & 0.11 & 0.18 \\
        \midrule
        {2} 
        & $F1_\text{Val}$ & 0.57 & 0.55 & 0.57 & 0.55 & 0.57 \\
        \cmidrule(lr){1-7}
        {\textcolor{softgreen}{$F1_\text{Cryptic}$}} 
        & Ours    & 0.41 & 0.38 & 0.35 & \textbf{0.33} & \textbf{0.34} \\
        & Rand5\% & 0.56 & 0.56 & 0.56 & 0.56 & 0.56 \\
        & RandAM  & \textbf{0.33} & \textbf{0.33} & \textbf{0.33} & \textbf{0.33} & \textbf{0.34} \\
        \cmidrule(lr){1-7}
        {\textcolor{softred}{$F1_\text{Mimetic}$}} 
        & Ours    & \textbf{0.27} & \textbf{0.25} & \textbf{0.24} & \textbf{0.23} & \textbf{0.20} \\
        & Rand5\% & 0.12 & 0.12 & 0.12 & 0.12 & 0.12 \\
        & RandAM  & 0.23 & 0.16 & 0.16 & 0.16 & 0.16 \\
        \midrule
        {3} 
        & $F1_\text{Val}$ & 0.52 & 0.51 & 0.51 & 0.53 & 0.51 \\
        \cmidrule(lr){1-7}
        {\textcolor{softgreen}{$F1_\text{Cryptic}$}} 
        & Ours    & 0.35 & 0.33 & 0.36 & 0.30 & \textbf{0.27} \\
        & Rand5\% & 0.50 & 0.50 & 0.50 & 0.50 & 0.50 \\
        & RandAM  & \textbf{0.27} & \textbf{0.27} & \textbf{0.27} & \textbf{0.27} & \textbf{0.27} \\
        \cmidrule(lr){1-7}
        {\textcolor{softred}{$F1_\text{Mimetic}$}} 
        & Ours    & \textbf{0.26} & \textbf{0.23} & \textbf{0.38} & \textbf{0.33} & \textbf{0.31} \\
        & Rand5\% & 0.12 & 0.12 & 0.12 & 0.12 & 0.12 \\
        & RandAM  & 0.21 & 0.21 & 0.21 & 0.21 & 0.21 \\
        \midrule
        {4} 
        & $F1_\text{Val}$ & 0.51 & 0.51 & 0.51 & 0.54 & 0.52 \\
        \cmidrule(lr){1-7}
        {\textcolor{softgreen}{$F1_\text{Cryptic}$}} 
        & Ours    & \textbf{0.30} & \textbf{0.38} & \textbf{0.37} & \textbf{0.41} & \textbf{0.34} \\
        & Rand5\% & 0.51 & 0.51 & 0.51 & 0.51 & 0.51 \\
        & RandAM  & 0.41 & 0.41 & 0.41 & \textbf{0.41} & 0.41 \\
        \cmidrule(lr){1-7}
        {\textcolor{softred}{$F1_\text{Mimetic}$}} 
        & Ours    & \textbf{0.30} & 0.21 & \textbf{0.33} & \textbf{0.33} & \textbf{0.32} \\
        & Rand5\% & 0.12 & 0.12 & 0.12 & 0.12 & 0.12 \\
        & RandAM  & 0.27 & \textbf{0.27} & 0.27 & 0.27 & 0.27 \\
        \bottomrule
    \end{tabular}

    \vspace{-3ex}
    
\end{table}

\section{Conclusion and Future Work} \label{sec:conclusion}

This paper introduced a novel IC camouflaging methodology that integrates ML-driven approaches and covert gates to promote hardware security and anti-reverse engineering. By bridging cryptic and mimetic cyber deception strategies, our approach achieved dual-layered camouflage, not only concealing the IP's functionality but also mimicking the appearance of the camouflaged IP as another misleading design. %The introduction of covert inverters, buffers, and universal transmitters further amplifies the robustness of our camouflaging technique, complicating adversarial efforts to decode true circuit functionalities.
Experimental validation demonstrated the effectiveness of the proposed method in achieving SAT resistance with low structural and performance overhead. Additionally, the methodology proved more resilient against AI-enhanced reverse engineering attacks than existing approaches.
These findings represent a significant advancement, setting a new standard for IC camouflaging by incorporating cyber deception principles. %This dual-layered defense mechanism provides a novel solution to safeguard critical systems, contributing to the community's broader goal of securing ICs in the face of evolving adversarial threats. 
Besides the exploration of promising new applications, our future work will focus on expanding the proposed methodology to generic process design kits or PDKs (as opposed to AIGs), investigating alternative AI techniques, and evaluating the effectiveness of fault and side-channel attacks.

\section*{Acknowledgments}
% \textit{Removed for double-blind review.}
The authors would like to thank Charles Kamhoua, Frederica Nelson, and Gregory Shearer of the Army Research Lab (ARL) for their encouragement and inspiration for this research. Also, this work has been supported in part by the US Army Research Office (ARO) under award \# W911NF-19-1-0102 and in part by the Department of Defense through the Science, Mathematics, and Research for Transformation (SMART) Scholarship-for-Service Program.

%\newpage
\bibliographystyle{IEEEtran}
\bibliography{references.bib}

\end{document}